\documentclass{article}
\usepackage{graphicx}

\begin{document}                                 
\noindent {\Large Low temperature specific heat of glasses : A non-extensive  approach}
\vskip1.5cm
\noindent {\bf Ashok Razdan}

\noindent {\bf Nuclear Research Laboratory }

\noindent {\bf Bhabha Atomic Research Centre }

\noindent {\bf Trombay, Mumbai- 400085 }
\vskip 1.5cm
\noindent {\bf Abstract :}

Specific heat  is calculated using Tsallis Statistics.
It is observed that it is possible to explain some low temperature specific heat 
properties  of glasses using non-extensive approach. A similarity between temperature dependence of
non-extensive specific heat and fractal specific heat 
is also discussed.

\noindent {\bf Motivation:}

Non-extensive statistics is being increasingly used to explain
anomalous behaviour observed in the properties of various physical 
system.    Tsallis statistics has been used to study
physical systems /phenomena which include turbulence in plasma [1], Cosmic ray background radiation [2], self gravitating systems  [3], econo-physics[4], electron -positron annihilation [5],classical and quantum chaos [6], linear response 
theory [7], Levy type anomalous super diffusion [8], thermalization of 
electron - phonon systems [9], low dimensional dissipative systems [10] etc. 
It has been shown that non-extensive features get manifested in those systems 
which have long range forces, long memory effects or in those systems which 
evolve in (non Euclidean like space-time) fractal space time [11 and reference therein]. 
Apart from other applications it has  been suggested [11] that non-extensive
statistics can be applied to complex systems like glassy materials and fractal /
multi-fractal or unconventional structures also.
Anomalous low temperature specific heat results in glasses  
have motivated us to use non-extensive statistics. We will also compare 
specific heat using Tsallis statistics with specific heat of
a fractal.

\noindent{\bf Low temperature specific heat:}

Specific heat depends on density of states $g(\omega)$
\begin{equation}
C_p= \frac{{k_B}^2}{\hbar} \int_{0}^{\infty} g(\frac{k_B T}{\hbar} x) \frac{ x^2 e^{x}}{(e^{x}-1)^2} dx
\end{equation}

where $x$ =$\frac{\hbar \omega}{k_B T}$.
Above equation has been obtained using Boltzmann Gibbs statistical mechanics.
Glasses at low temperature (below 2K) show
quasi-linear behaviour i.e. approximate linear dependence on temperature T. Again  glassy systems  
do not follow $T^3$ dependence even above 2K as in 
some cases  maximum in $\frac{C_p}{T^3}$  is observed. Above properties are  universal 
features of glassy systems [12,13].
Another interesting feature is $\frac{C_p}{T}$ dependence on $T^2$ for crystals and 
glassy phase of the same systems [14].

\noindent {\bf Theory :}

Non-extensive statistics is based on two postulates [11 and reference therein].
First is the definition of non-extensive entropy
\begin{equation}
S_q  = \frac{ 1- P_i^{q}}{q -1}                              
\end{equation}

where q is 
the non-extensive entropic index and $P_i$ are the 
probabilities of the microscopic states with $\sum P_i$  = 1.

The second postulate is the definition of  energy $U_q$  = $\sum P_i^{q} E_i , $ 
where $E_i$ is the energy spectrum. 
As q $\rightarrow$1, $S_q$  = - $\sum p_i ln P_i $, which is the Boltzmann Gibbs Shannon 
entropy.

To derive non-extensive form of specific heat we have to use 
non-extensive quantum distribution function of bosons. It is very 
difficult to derive exact analytical expression for non-extensive distribution 
function. However, there are many studies which provide the approximate 
form of non-extensive distribution functions [2,15]. In the present paper we use 
dilute gas approximation (DGA) for boson distribution function. For DGA 
case, the average occupation number is given as [15]
\begin{equation}
< n_q > = \frac{1}{(1+(q-1)\beta (E_i -\mu)^{\frac{1}{q-1}}-1)}
\end{equation}
\begin{figure}
\begin{center}
\includegraphics[angle=270,width=7.0cm]{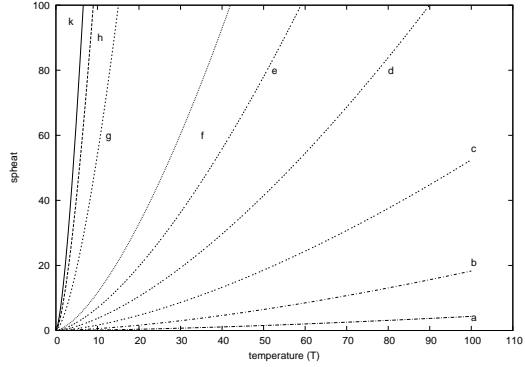} 
\caption{Non-extensive specific heat for various values of q; q=1.5(a),1.6(b),1.7(c),1.8(d),
1.9(e),2.0(f),2.5(g),3.0(h),3.5(k)}
\end{center}
\end{figure}
When dealing with system in contact with the heat bath at the temperature 
$\beta$, we have 
\begin{equation}
< n_q > =\frac{\hbar \omega_i}{(1+(q-1) \frac{\hbar\omega_i}{kT})^{\frac{1}{q-1}}-1}
\end{equation}
Averaged over non-extensive distribution, expectation value of energy is
\begin{equation}
< E_q >= \hbar \omega < n_q >
\end{equation}
and total vibrational energy is 
\begin{equation}
U_q = \int \hbar \omega < n_q > g(\omega)
\end{equation} 
from  which specific heat can be obtained  
\begin{equation}
C_p =\frac{\partial U_q}{\partial T}
\end{equation}
\begin{figure}
\includegraphics[angle=270,width=7.0cm]{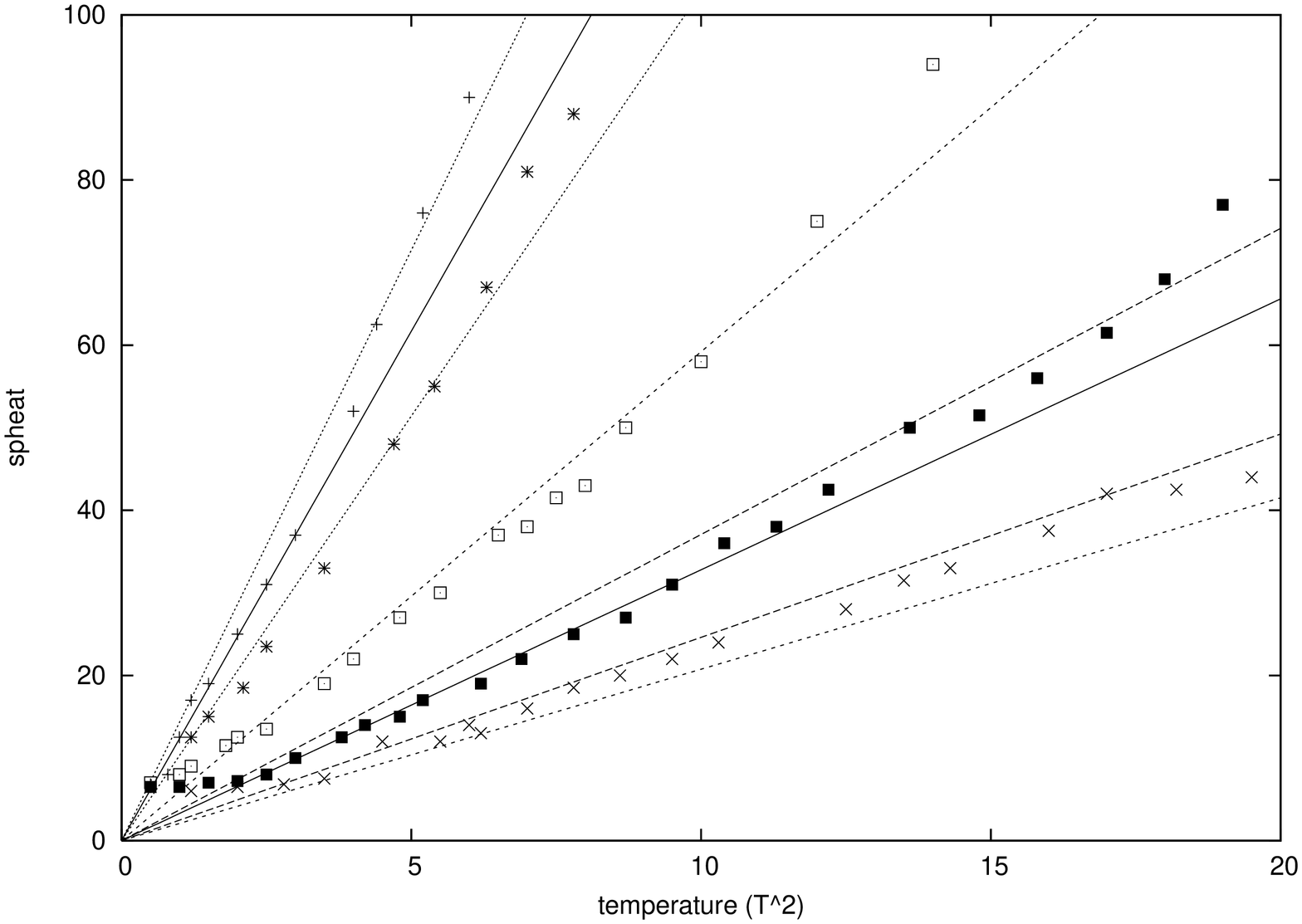} \includegraphics[angle=270,width=7.0cm]{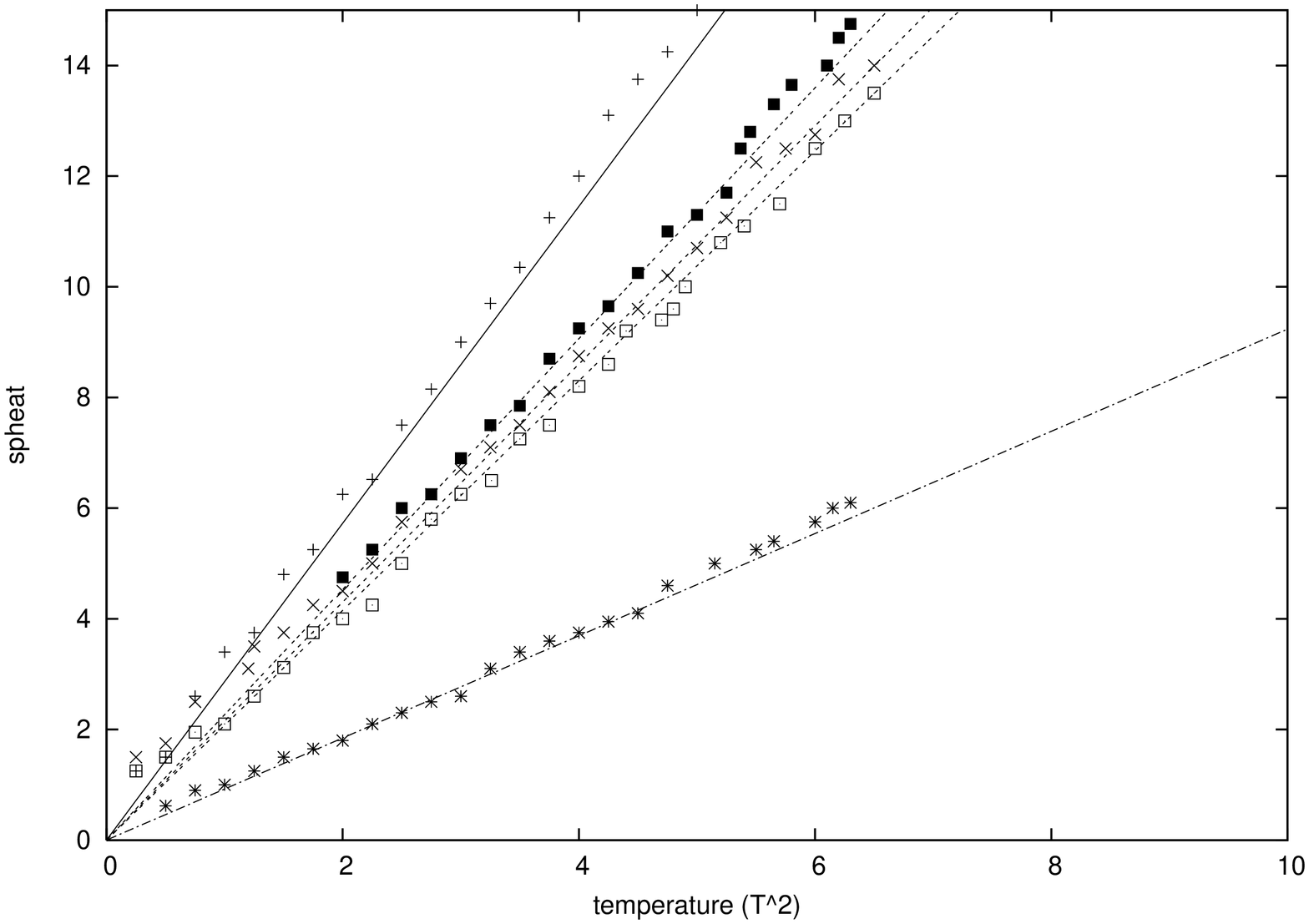}
\caption{Fig2a: Sp. heat data  corresponding to $(B_2 O_3)_{100-x} (Na_2 O)_x$ for different $\%$ of $Na_2 O$
[16,17,18] 
Fig2b: Sp. heat data corresponding to glassy alcohols [14,19]. In y-axis 
$\frac{c_p}{T}$ is plotted against $T^2$ in x-axis}

\end{figure}
For the case of Debye approximation    
$g(\omega) \propto (\omega)^2$.
Equation (7) represents specific heat using Tsallis statistics 
where $n_q$  is given by equation (4).

\noindent {\bf Results and Discussion :}

We have numerically solved equation (7) using equations (4) and (6), for  various values of q. 
It is very clear from figure 1 that quasi-linear behaviour or maxima in $\frac{c_p}{T^3}$ is not explained by including
non-extensive statistics. Various  curves in figure 1 have been obtained for fixed value of
Debye temperature ($\theta_D$=300 K). The shape and nature of  curves in figure 1 is decided by  value of
q.
A linear display of specific  heat versus $T^2$ dependence is followed in glasses as depicted in figure 2a and 
figure 2b.
In figure 2a we are plotting experimental data corresponding to $(B_2 O_3)_{100-x} (Na_2 O)_{x}$ taken from
reference [16,17,18]. The data with (+) sign corresponds to 0 $\%$ of $Na_2 O$, (*) corresponds to 1 $\%$ of $Na_2 O$,
hollow square corresponds to 6 $\%$, filled square corresponds to 16 $\%$ and crosses (x) corresponds to
25 $\%$ of $Na_2 O$ respectively. 
The fitted  curves correspond to different values of q from (+)onwards are q=5.5, 5.0, 4.5, 3.5,3.0,2.9,
2.7,2.6 respectively. 
The slope of experimental data in figure 2a is strongly dependent on presence, absence or excess of
$Na_2 O$ in glassy system. Non-extensive specific heat seems to explain data in figure 2a for different values of q. Again in figure
2b specific heat data of various glassy systems show strong dependence on slope and can be explained by non-extensive 
approach. 
The experimental data in fig2b corresponds to glassy alcohols like 2-proponal (+), filled square to D-ethonal,
cross (x) to D-ethonal (OG,  orientational glass), hollow square to 1-proponal and star (*) to glycerol respectively.
The corresponding fitted q-values are q=2.8,2.65,2.62,2.60 and 2.25 respectively. The experimental data has
been taken from reference [19,22].
It is interesting to mention here that  a strong slope dependence of specific heat on $T^2$ has been used
to [14] depict two different phases  
of crystal and glasses with two different Debye temperatures.  

Low temperature specific heat studies in silica aerogels and other disordered systems do not follow Debye model
and concepts of fractals have been used to explain such results [20] . For a fractal approach density of states 
$g(\omega) \propto \omega^{\tilde{d}}$ where $\tilde{d}$ is spectral dimension.

\begin{figure}
\begin{center}
\includegraphics[angle=270,width=7.0cm]{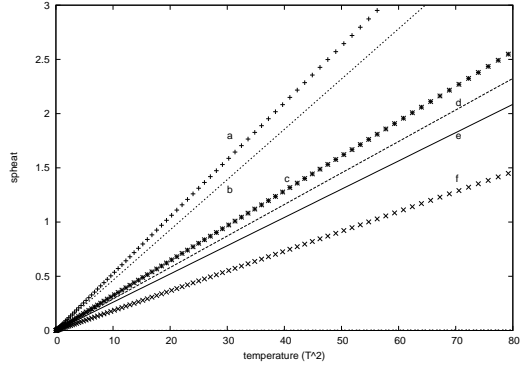}
\caption{Comparison of non-extensive specific heat with fractal specific heat. In y-axis
$\frac{c_p}{T}$ is plotted against $T^2$ in x-axis}
\end{center}
\end{figure}

We have plotted  proportional temperature dependence of
specific heat of a fractal in figure 3. Curves b,d and e correspond to $\tilde{d}$=3.8,2.8 and 1.8
whereas curves a,b and f correspond to q=1.7,1.65 and 1.6 respectively. 
It is evident from figure 3 that there is lot of similarity between fractal and
non-extensive specific heat.
The similarity between these curves is expected  because 
non-extensive statistics corresponds to non-Euclidean space time which 
is fractal in nature. The similarity between fractal and nonextensive results were also 
reported for Lamb Mossbauer factor [21]. It has been shown that magnetic properties
of manganites [23] 
and their corresponding phase diagrams [24] can be explained in the framework of Tsallis statistics.

Some of the anomalous properties  have been explained by using soft potential model which is
an extension of tunnelling model in which group of atoms can tunnel between same energy
configurations [14]. Recently it has been reported that 
quasi-linear behaviour  or  maximum in
specific heat data of various incommensurate phases
can be explained  [25] by considering effect of phason damping.

\noindent {\bf References: }
\begin{enumerate}
\item B.M.Boghosian, Phys. Rev. E 53(1996)4745
\item C.Tsallis, F.C. Sa Barreto and E.D.Loh, Phys. Rev. E 52(1995) 1447
\item V.H.Hamity and D.E.Barraco, Phys. Rev. Lett. 76(1996)4664
\item C.Tsallis, C.Anteneodo, L.Borland and R.Osorio, Cond-mat/0301307
\item I.Bediaga, E.M.F.Curado and J. Miranda, Physica A 286(2000)156
\item C.Tsallis, A.R.Plastino and W. -M.Zheng, Choas,Solitons and Fractals 8(1997)885,\\
      Y.Weinstein, S.Lloyd and C.Tsallis, Phys. Rev. Lett. 89(2002)214101
\item A.K.Rajagopal, Phys. Rev. Lett. 76(1996) 3496
\item C.Tsallis, S.V.F. Levy, A.M.C.Souza and R.Maynard, Phys. Rev. Lett. 75(1995)3589
\item I.Koponen, Phys. Rev. E 55(1997)7759
\item M.L.Lyra and C.Tsallis, Phys. Rev. Lett. 80(1998)53
\item C.Tsallis, Physica A 221(1995)277-290,\\  
      C.Tsallis, R.S.Mendes, A.R.Plastino, Physica A 261(1998)534
\item W.A.Phillips, editor Amorphous Solids: Low Temperature Properties, Springer Berlin 1981
\item W.A.Phillips, Rep. Prog. Phys. 50(1987)1657
\item M.A.Ramos, C.Talon and S.Vieira, Phys. Rev. B 66(2002)01 2201
\item Q.A.Wang and A.Le Mehaute, Phys. Lett. A 242(1998)301
\item E.S.Pinango, M.A.Ramos, R.Villar and S.Vieira, in  Basic Features of glassy State; Edited by 
J.Colmenero and A.Alegria (Singapore: World Scientific),pp. 509-513
\item M.A.Ramos and U.Buchenau 1998, Tunnelling Systems in Amorphous and Crystalline Solids, Edited by
P.Esquinazi (Berlin:Springer) chapter 9,pp-527-569
\item M.A.Ramos Cond-mat/0405055
\item M.A.Ramos, C.Talon and S.Vieira Cond-mat/0201560 v1
\item A.M.de Goer, R.Calemczuk, B.Slace, et al Phys. Rev. B 40(1989)8327
\item A.Razdan, Phys. Lett. A 321(2004)190
\item M.A.Ramos, C.Talon, R. J. Jimenez-Rioboo and S. Vieira, J. Phys: Condens. Matter 15(2003)S1007
\item V.S.Amaral, E.K.Lenzi and I.S. Oliveira, Phys. Rev. B 66(2002)134417
\item M.S.Reis, V.S.Amaral, J.P. Araujo and I.S. Oliveira, Phys. Rev. B 68(2004)014404
\item A.Cano and A.P.Levanyuk, Phys. Rev. Lett. 93(2004)245902
\end{enumerate}
\end{document}